\documentclass[twocolumn]{bmcart}

\usepackage{multirow}
\RequirePackage{url,hyperref}
\usepackage[utf8]{inputenc} 

\def\includegraphics{}

\usepackage{amssymb}
\usepackage{setspace}
\usepackage{graphics,epsfig,cite,subfigure,amsmath,amsthm}
\usepackage[all]{xy}

\startlocaldefs
\endlocaldefs

\newtheorem{theorem}{Method}[section]   

\theoremstyle{definition}
\newtheorem{definition}[theorem]{Definition}   

\theoremstyle{remark}
\newtheorem{example}[theorem]{Example}        

\begin{document}

\begin{frontmatter}

\begin{fmbox}
\dochead{Research}


\title{Identification of control targets in Boolean molecular network models via computational algebra}


\author[
   addressref={aff1},                   
   corref={aff1},                       
   email={murrugarra@uky.edu}   
]{\inits{DM}\fnm{David} \snm{Murrugarra}}
\author[
   addressref={aff2},
   email={avelizcuba1@udayton.edu}
]{\inits{ESD}\fnm{Alan} \snm{Veliz-Cuba}}
\author[
   addressref={aff3},
   email={boaguilar@gmail.com}
]{\inits{ESD}\fnm{Boris} \snm{Aguilar}}
\author[
   addressref={aff4,aff5},
   email={laubenbacher@uchc.edu}
]{\inits{ESD}\fnm{Reinhard} \snm{Laubenbacher}}

\address[id=aff1]{
  \orgname{Department of Mathematics, University of Kentucky}, 
  \city{Lexington, KY 40506-0027},                              
  \cny{USA}                                    
}
\address[id=aff2]{%
  \orgname{Department of Mathematics, University of Dayton},
  \city{Dayton, OH 45469},
  \cny{USA}
}
\address[id=aff3]{%
  \orgname{Institute for Systems Biology},
  \city{Seattle, WA 98109-5263},
  \cny{USA}
}
\address[id=aff4]{%
  \orgname{Center for Quantitative Medicine, University of Connecticut Health Center},
  \city{Farmington, CT 06030-6033},
  \cny{USA}
}
\address[id=aff5]{%
  \orgname{Jackson Laboratory for Genomic Medicine},
  \city{Farmington, CT 06030},
  \cny{USA}
}

\begin{artnotes}
\end{artnotes}


\begin{abstractbox}
\begin{abstract} 
\parttitle{Background}
Many problems in biomedicine and other areas of the life sciences can be characterized as control problems, with the goal of finding strategies to change a disease or otherwise undesirable state of a biological system into another, more desirable, state through an intervention, such as a drug or other therapeutic treatment. The identification of such strategies is typically based on a mathematical model of the process to be altered through targeted control inputs. This paper focuses on processes at the molecular level that determine the state of an individual cell, involving signaling or gene regulation. The mathematical model type considered is that of Boolean networks. The potential control targets can be represented by a set of nodes and edges that can be manipulated to produce a desired effect on the system.
\parttitle{Results}
This paper presents a method for the identification of potential intervention targets in Boolean molecular network models using algebraic techniques. The approach exploits an algebraic representation of Boolean networks to encode the control candidates in the network wiring diagram as the solutions of a system of polynomials equations, and then uses computational algebra techniques to find such controllers. The control methods in this paper are validated through the identification of combinatorial interventions in the signaling pathways of previously reported control targets in two well studied systems, a p53-mdm2 network and a blood T cell lymphocyte granular leukemia survival signaling network. Supplementary data is available online and our code in Macaulay2 and Matlab are available via \href{http://www.ms.uky.edu/~dmu228/ControlAlg}{http://www.ms.uky.edu/$\sim$dmu228/ControlAlg}.
\parttitle{Conclusions}
This paper presents a novel method for the identification of intervention targets in Boolean network models. The results in this paper show that the proposed methods are useful and efficient for moderately large networks.
\end{abstract}

\end{abstractbox}
\end{fmbox}

\end{frontmatter}
\section{Background}
The dynamics of gene regulatory networks (GRNs) have been studied using different modeling frameworks,
with the goal of building computational models of GRNs to get new insights into important cellular processes, e.g.,
the cell cycle,~\cite{Tyson:2001aa,Li:2004aa}, oscillations in the \textit{p53-mdm2} system,~\cite{Geva-Zatorsky:2006aa,Batchelor:2009aa,Choi2012},
the phage-lambda system,~\cite{DBLP:journals/ploscb/JohW11,PMID:20478257,Murrugarra2012},
or the T cell large granular lymphocyte (T-LGL) leukemia network,~\cite{Zhang:2008aa,Saadatpour:2011aa}.
A generally difficult problem is to design control policies to achieve desired dynamics in GRNs.
This is particularly important in the control of cancer cells,~\cite{Choi2012,Wang:2013aa,Lee:2012aa,Wang:2014qy,Erler:2012zr}
and cell fate reprogramming,~\cite{PMID:16904174,Young:2011aa}.
Thus, developing tools to control mathematical models of GRNs are key to the design of experimental control policies.

There is a rich theory for the control of continuous models such as systems of differential equations,~\cite{Iglesias:2010aa,Shin:2010aa,Cornelius:2013aa,Motter:2015kx}.
Discrete models such as Boolean networks (BN) have been proposed to study GRNs. In a BN, the genes of a GRN are represented by a set of nodes that can take on only two possible states (ON or OFF).
Time is discrete, and the state of a gene at the next time step is determined by a Boolean function over a subset of nodes of the BN. The dependence of a gene on the state of another gene is graphically represented by a directed edge. BN models are widely used in molecular and systems biology to capture coarse-grained dynamics of a variety of regulatory networks and have been shown to provide a good approximation of the dynamics of continuous processes~\cite{Albert:2003aa,Kauffman2003,DBLP:journals/ploscb/Saez-RodriguezSLHBAHWGKS07,Balleza:2008aa,Davidich:2008aa,DBLP:journals/jcb/Veliz-CubaS11,Saadatpour:2011aa,Murrugarra2012,Veliz:BNODE2012,Veliz:BNODE2014}.
However, control methods for discrete models are still in their infancy, compared to the theory for continuous models.

In this work, we propose a framework to study the control of BNs. Therapeutic interventions are modeled by manipulating the wiring diagram of a BN accordingly. 
For example, a gene knockout is modeled by fixing the value of one node of the BN to OFF. 
Similarly, deleting directed edges of a BN models the action of a drug that inactivates the interaction between two gene products.
The control problem consists of finding the sequence of actions (node and edge deletions) to get  the BN out of an undesirable state,
and drive it  towards a desirable state. Undesirable states may represent a disease condition of a cell such as, for example, a highly
proliferative state of a cancer cell, and a desirable state may represent cell death. Thus, a therapeutic intervention could aim at inducing the death of aberrant tumor cells.

Several approaches to address this problem have been used in recent years. 
For example, the optimal control techniques developed in~\cite{yousefi2012,Yousefi15072013,6557489,Yousefi:2014aa} assume a set of control nodes to derive a control policy that minimizes the likelihood of transitioning into an undesirable state in a stochastic context. Other approaches for the identification of intervention targets include the use of stable motifs in the network
to induce the system into a desirable state~\cite{Zanudo:2015aa}. In~\cite{Qiu:2014aa},
integer programming was used to find a set of nodes to control the states of BNs representing tumor and normal cells.
Optimization techniques were used in \cite{Shmulevich01102002} to determine possible node manipulations for BNs.
There are also search algorithms based on genetic and greedy algorithms described in \cite{DBLP:journals/bioinformatics/XiaoD07,Vera-Licona:2013aa,PMID:25433558}. For continuous models, a related control approach is given in~\cite{Cornelius:2013aa}.

The idea behind our approach is to encode the problem of finding control candidates
as a problem of solving a system of polynomial equations. Then we can use computational algebra techniques to solve the system. This approach takes advantage of the rich algorithmic theory of computer algebra (e.g. Gr\"obner basis techniques) and the theoretical
foundations of algebraic geometry with software available for computations (e.g., Macaulay2~\cite{macaulay2}).
The output of our method is a set of candidate actions with the capacity to induce the GRN towards desirable states.
The method also has the ability of identifying combinatorial interventions in the network which are sometimes more effective as will be shown in the results section.
The algebraic view of discrete models has been previously used for the development of computational tools to determine the attractors of
BNs~\cite{Veliz-Cuba:2014aa,Veliz-Cuba:2010aa,Hinkelmann:2011aa,Veliz:redbn}, and also for inferring network structure from dynamics~\cite{Curto:2013,Veliz:Rev_Eng,jarrah2007,Laubenbacher:2004aa}. 
\section{Methods}
\subsection{Boolean Networks}
A Boolean network is a dynamical system that is discrete in time as well as 
in variable states. More formally,
consider a collection $x_1, \ldots , x_n$ of
variables that take values in the binary set $\{0,1\}$.
Then a Boolean network in the variables $x_1, \ldots , x_n$ is a function 
\begin{displaymath}
\mathbf{ F} = (f_1,\dots,f_n):\{0,1\}^n\rightarrow \{0,1\}^n
\end{displaymath}   
where each coordinate function $f_i:\{0,1\}^n\rightarrow \{0,1\}$ is a Boolean function on a subset of $\{x_1,\dots,x_n\}$
which represents how the future value of the $i$-th variable depends on the present values of the other variables. 
\begin{example}\label{eg:toy}
For concreteness, we illustrate the definitions using the following toy network
\begin{displaymath}
\begin{array}{lll}
f_1  = \neg x_3 \wedge \neg x_5, \quad &  f_2  = \neg x_1 \vee x_4,  \quad &
f_3  = \neg x_2, \\
f_4  = x_3, & f_5  = \neg x_4,
\end{array}
\end{displaymath}   
where $\wedge$, $\vee$, and $\neg$ are the AND, OR, and NOT logical operators, respectively. In the context of modeling biological systems, $\wedge$ corresponds to activation by the combination of regulators (all regulators are necessary), $\vee$ corresponds to independent activation (one regulator is sufficient), and $\neg$ corresponds to negative regulation.

\end{example}
\begin{figure}[!h]
\begin{center}
\includegraphics[width=3in]{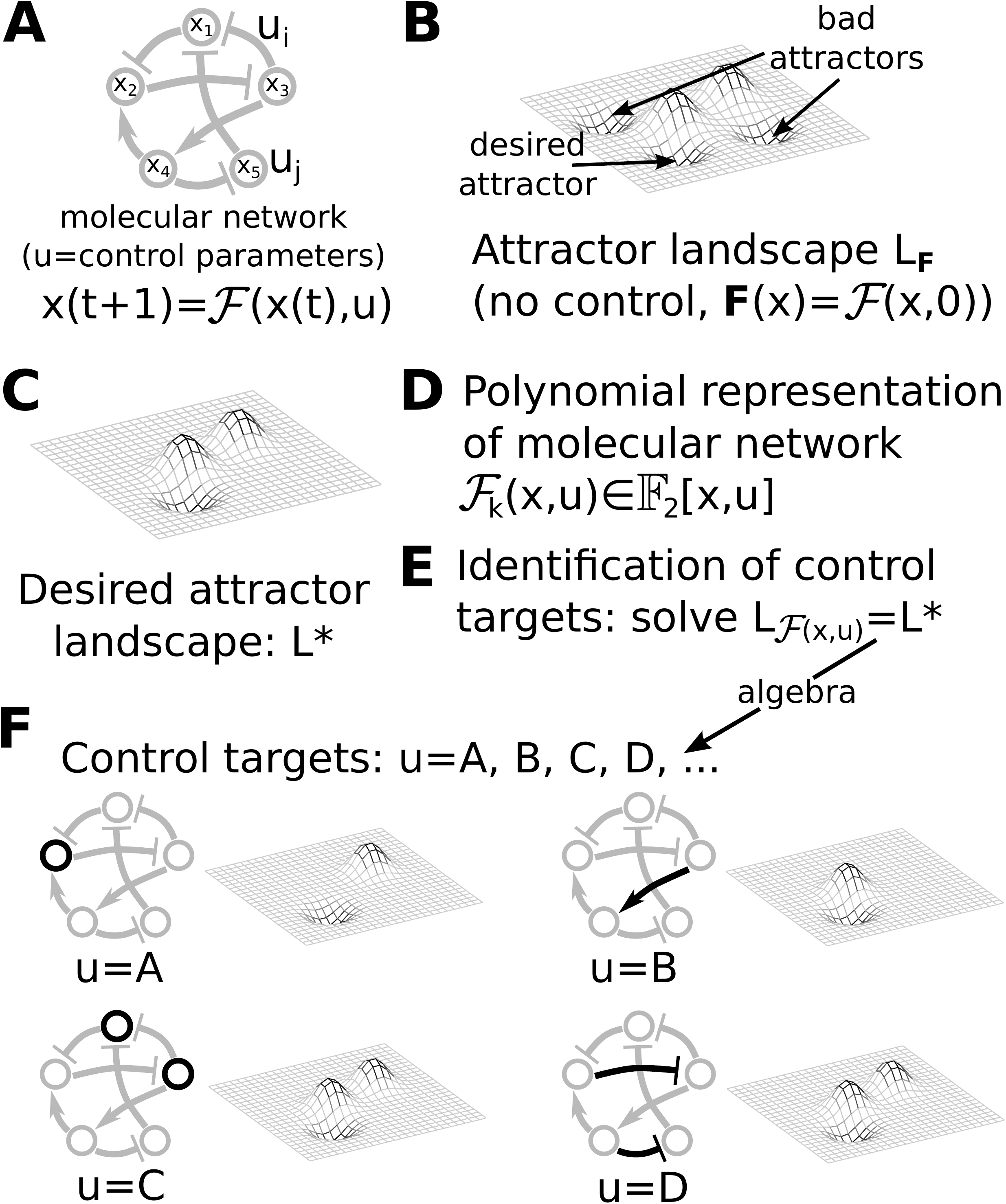}
\caption{Description of the algebraic approach of identification of control targets.
\textbf{A.} A BN model of a molecular network. The control variables are the entries of $u$. 
\textbf{B.} In the absence of control policies ($u=0$), the attractor landscape ($L_F$) can have undesirable attractors. 
\textbf{C.} The goal is to choose control values that give a desired attractor landscape, $L^*$.
\textbf{D.} To use the algebraic approach we first find the polynomial representation of the BN (see Section \ref{sec:methods_defcontrol}).
\textbf{E.} The next step is to set up the desired attractor landscape as a system of equations that the BN has to satisfy, $L_{\mathcal{F}(x,u)}=L^*$ (see Section \ref{sec:methods_equations}). 
\textbf{F.} Solving the equation $L_{\mathcal{F}(x,u)}=L^*$ for $u$ will provide the control values to achieve the desired landscape (see Section \ref{sec:methods_solving}). This approach not only finds individual control policies ($u=A$, single node; $u=B$, single edge), but also \emph{combinatorial} control policies ($u=C$, two nodes; $u=D$, two edges). In a combinatorial control policy, the desired attractor landscape is achieved by the combination of two or more entries of $u$.
}
\label{fig:summary}
\end{center}
\end{figure}  

Given a Boolean network $\mathbf{ F} = (f_1,\dots,f_n)$, a directed graph $\mathcal{W}$ with $n$ nodes 
$x_1, \ldots , x_n$ is associated with $\mathbf{ F}$. There is a directed edge in $\mathcal{W}$ from 
$x_j$ to $x_i$ if $x_j$ appears in $f_i$. 
Notice that the presence of the interaction $x_j\rightarrow x_i$ implies that the regulatory function $f_i$ depends on $x_j$, say $f_i(x_{k_1},\dots,x_j,\dots,x_{k_m})$ with $x_j\in\{x_{k_1},\dots,x_{k_m}\}$.
In the context of a molecular network model, $\mathcal{W}$ represents the wiring diagram of the network.

\textit{Example }\ref{eg:toy} (cont.)  The wiring diagram of the toy network is shown in Figure~\ref{fig:summary}A. 

If the set
$\{0,1\}$ is given the structure of a finite field with standard addition and multiplication, $\mathbb F_2 = \{0,1\}$, then the functions 
$f_i:\mathbb F_2^n\rightarrow \mathbb F_2$ can be represented as polynomials over $\mathbb F_2$, and the dynamical system
$\mathbf{ F} = (f_1,\dots,f_n):\mathbb F_2^n\rightarrow \mathbb F_2^n$
becomes a polynomial dynamical system, see~\cite{Veliz-Cuba:2010aa}, which gives access to
a range of mathematical tools for the analysis of $\mathbf{F}$.

\textit{Example }\ref{eg:toy} (cont.)
The rules to transform Boolean functions into polynomials in $\mathbb{F}_2[x_1,\ldots,x_n]$ are given by $a\wedge b=ab$, $a\vee b=a+b+ab$, and $\neg a=1+a$, where the operations are computed modulo 2. For the toy network we obtain
\begin{displaymath}
\begin{array}{l}
f_1  = (1+ x_3)(1+x_5)=1+x_3+x_5+x_3x_5,\\
f_2  = (1+ x_1)+ x_4 +(1+ x_1)x_4 = 1+x_1+x_1x_4,\\
f_3  = 1+ x_2, \\
f_4  = x_3,\\
f_5  = 1+ x_4. 
\end{array}
\end{displaymath}

The dynamical properties of a Boolean network are given by the difference equation $x(t+1)=\mathbf{ F}(x(t))$; that is, the dynamics is generated by iteration of $\mathbf{ F}$. More precisely,
the dynamics of $\mathbf{ F}$ is given by the state space graph $S$, defined as the graph with vertices in
$\mathbb F_2^n=\{0,1\}^n$, which has an edge from $x\in \mathbb F_2^n$ to $y\in \mathbb F_2^n$ if and only 
if $y = \mathbf{ F}(x)$. The states $x\in \mathbb F_2^n$ where the system will get stabilized are of particular importance. These special points of the state space are called attractors of a
Boolean network and these may include steady states (fixed points), where $\mathbf{ F}(x) = x$, and cycles, where $\mathbf{ F}^k(x) = x$ for some integer $k>1$.
Attractors in Boolean network modeling might represent cell types~\cite{Kauffman:1969aa} or cellular states such as apoptosis, proliferation, or cell senescence~\cite{Huang:1999aa,DBLP:books/daglib/0024105}. Identifying the attractors of a system is an important step towards the control of that system and can be done using tools from computational algebra \cite{Veliz-Cuba:2010aa,Veliz-Cuba:2014aa}.

\textit{Example }\ref{eg:toy} (cont.) The steady states of the toy network are found by solving the system of equations $f_i=x_i$, $i=1,\ldots,5$. This means we want to find the roots of $g_i=0$, where $g_i=f_i-x_i$. This gives the system of equations
\begin{equation}
\label{Eq:toy_poly_sys}
\begin{array}{l}
g_1 = 1+x_3+x_5+x_3x_5 + x_1=0,\\
g_2 = 1+x_1+x_1x_4 + x_2=0,\\
g_3 = 1+ x_2 + x_3=0, \\
g_4 = x_3 + x_4=0,\\
g_5 = 1+ x_4 + x_5=0. 
\end{array}
\end{equation}
Note that, since the system is not linear, we cannot make use of tools such as Gaussian elimination. Computational algebra allows us to solve such systems by encoding the solutions as an algebraic object called an ideal of polynomials, $I=\langle g_1,g_2,g_3,g_4,g_5 \rangle$, and then finding an equivalent, but simpler representation. More precisely, for the system in Equation~\ref{Eq:toy_poly_sys} we can find its Gr\"{o}bner basis~\cite{cox1998using} and obtain
\begin{displaymath}
I
= \langle g_1,g_2,g_3,g_4,g_5 \rangle 
= \langle x_5+1, x_4, x_3, x_2+1 , x_1 \rangle.
\end{displaymath} 

This means that the original system has the same solutions as the simpler system
\begin{displaymath}
\begin{array}{l}
 x_1 = 0,
 x_2+1 =0,
 x_3 = 0,
 x_4 =0,
 x_5+1 =0,
\end{array}
\end{displaymath} 
which is easily solved to obtain $x=01001$ as a steady state of the toy network (parentheses omitted for brevity).

\subsection{Control Actions: edge and node manipulations}
\label{sec:methods_defcontrol}
This paper considers two types of control actions: 1. Deletion of edges and 2. Deletion (or constant expression) of nodes.
The motivation for considering these actions is that they represent the common interventions that prevent a regulation from happening.
For instance, an edge deletion can be achieved by the use of therapeutic drugs that target specific gene interactions and node deletions represent the blocking of effects of products of genes associated to these nodes; see~\cite{Choi2012}. 

A schematic of our approach is given in Figure~\ref{fig:summary} where the dynamics of a Boolean network can be manipulated by a set of controls consisting of deletion or constant expressions of edges and nodes. Below we define and explain all the steps in more detail.

A Boolean network with control is given by a function $\mathcal{ F}:\mathbb{F}_2^n\times U\rightarrow \mathbb{F}_2^n$,
where $U$ is a set that denotes all possible control inputs (Figure \ref{fig:summary}A). Given a control $u$ in $U$, the dynamics are given by $x(t+1)=\mathcal{F}(x(t),u)$.

We consider a BN $\mathbf{F}=(f_1,\ldots,f_n):\mathbb{F}_2^n\rightarrow \mathbb{F}_2^n$ and show how to encode edge and node controls by $\mathcal{F}:\mathbb{F}_2^n\times U \rightarrow \mathbb{F}_2^n$, such that $\mathcal{F}(x,0)=\mathbf{F}(x)$. That is, the case of no control coincides with the original BN. We remark that these control types can be combined, but for clarity we present them separately.
\begin{definition}[Edge Control]\label{def:edge_del}
Consider the edge $x_i\rightarrow x_j$ in the wiring diagram $\mathcal{W}$. 
The function
\begin{equation}
\label{edge_del_def}
\mathcal{F}_j(x,u_{i,j}) := f_j(x_1,\dots,(u_{i,j}+1)x_i,\dots,x_n)
\end{equation}
encodes the control of the edge $x_i\rightarrow x_j$, since for each possible value of $u_{i,j}\in \mathbb{F}_2$ we have the following control settings: 
\begin{itemize}
  \item If $u_{i,j}=0$, $\mathcal{F}_j(x,0) = f_j(x_1,\dots,x_i,\dots,x_n)$. That is, the control is not active.
  \item If $u_{i,j}=1$, $\mathcal{F}_j(x,1) = f_j(x_1,\dots,x_i=0,\dots,x_n)$. In this case, the control is active, and the action represents the removal of the edge $x_i\rightarrow x_j$.
\end{itemize}
Definition~\ref{def:edge_del} can easily be extended for the control of many edges, so that we obtain $\mathcal{F}:\mathbb{F}_2^n\times \mathbb{F}_2^e \rightarrow \mathbb{F}_2^n$, where $e$ is the number of edges in the wiring diagram. Each coordinate, $u_{i,j}$, of $u$ in $\mathcal{F}(x,u)$ encodes the control of an edge $x_i\rightarrow x_j$.
\end{definition}

\textit{Example }\ref{eg:toy} (cont.) Incorporating edge control in the toy network results in 
\begin{displaymath}
\begin{array}{l}
\mathcal{F}_1  = 1+(u_{3,1}+1)x_3+(u_{5,1}+1)x_5+\\  
\quad \quad \quad (u_{3,1}+1)x_3(u_{5,1}+1)x_5,\\
\mathcal{F}_2  =  1+(u_{1,2}+1)x_1+(u_{1,2}+1)x_1(u_{4,2}+1)x_4,\\
\mathcal{F}_3  = 1+ (u_{2,3}+1)x_2, \\
\mathcal{F}_4  = (u_{3,4}+1)x_3,\\
\mathcal{F}_5  = 1+ (u_{4,5}+1)x_4. 
\end{array}
\end{displaymath} 

\begin{definition}[Node Control]\label{def:node_del}
Consider the node $x_i$ in the wiring diagram $\mathcal{W}$. The function
\begin{equation}
\label{node_del_def}
\mathcal{F}_j(x,u^{-}_i,u^{+}_i) := (u^{-}_i+u^{+}_i+1)f_j(x) + u^{+}_i
\end{equation}
encodes the control (knock-out or constant expression) of the node $x_i$, since for each possible value of $(u^{-}_i,u^{+}_i)\in \mathbb{F}_2^2$ we have the following control settings:  
\begin{itemize}
  \item For $u^{-}_i=0, u^{+}_i=0$, $\mathcal{F}_j(x,0,0) = f_j(x)$. That is, the control is not active.
  \item For $u^{-}_i=1, u^{+}_i=0$, $\mathcal{F}_j(x,1,0)  = 0$. This action represents the knock out of the node $x_j$.
  \item For $u^{-}_i=0, u^{+}_i=1$, $\mathcal{F}_j(x,0,1)  = 1$. This action represents the constant expression of the node $x_j$.
  \item For $u^{-}_i=1, u^{+}_i=1$, $\mathcal{F}_j(x,1,1) =  f_j(x_{t_1},\dots,x_{t_m})+1$. This action changes the Boolean function to its negative value and might not be a relevant case of control.
\end{itemize}
\end{definition}

\textit{Example }\ref{eg:toy} (cont.) Incorporating node control in the toy network results in 
\begin{displaymath}
\begin{array}{l}
\mathcal{F}_1  = (u_1^-+u_1^+ + 1)(1+x_3+x_5+x_3x_5) + u_1^+,\\
\mathcal{F}_2  = (u_2^-+u_2^+ + 1)(1+x_1+x_1x_4) + u_2^+,\\
\mathcal{F}_3  = (u_3^-+u_3^+ + 1)(1+ x_2) + u_3^+, \\
\mathcal{F}_4  = (u_4^-+u_4^+ + 1)x_3 + u_4^+,\\
\mathcal{F}_5  = (u_5^-+u_5^+ + 1)(1+ x_4) + u_5^+. 
\end{array}
\end{displaymath} 

\subsection{Control targets in Boolean networks.} 
\label{sec:methods_equations}
We consider a BN with control $\mathcal{F}:\mathbb F_2^n\times U \rightarrow \mathbb F_2^n$, and denote by $\mathbf{F}$ the BN with no control ($\mathcal{ F}(x,0) = \mathbf{F}(x)$). We remark that in each case of interest, both edge and node control could be analyzed simultaneously.

\subsubsection{Generating new steady states}
\label{sec:find_edge_del}
Suppose that $\textbf{y}_0 = (y_{01},\dots,y_{0n})\in \mathbb F_2^n$ is a desirable cell state (for instance, it could represent the state of cell senescence) but is not a fixed point, i.e., $\mathbf{ F}(\textbf{y}_0)\neq \textbf{y}_0$. The problem is then to choose a control $u$ such that  $\mathcal{F}(\mathbf{y}_0,u)= \mathbf{y}_0$. We now show how this can be achieved in our framework.

After encoding our BN with control as a polynomial system $\mathcal{F}_j(x,u)\in\mathbb{F}_2[x,u]$ (see Section \ref{sec:methods_defcontrol}), we consider the system of polynomial equations in the $u$ parameters:
\begin{equation}
\label{eq:forceFP}
\mathcal{F}_j(\textbf{y}_0,u)- y_{0j}=0 , j = 1,\dots,m.
\end{equation} 

\textit{Example }\ref{eg:toy} (cont.) Here we consider a toy network and assume we are interested in controlling edges to make $\textbf{y}_0=01111$ a steady state. For simplicity we only consider control using the edges $x_1\rightarrow x_2, x_4\rightarrow x_2, x_2\rightarrow x_3, x_4\rightarrow x_5$. The network is 
\begin{displaymath}
\begin{array}{l}
\mathcal{F}_1  = 1+x_3+x_5+x_3x_5,\\
\mathcal{F}_2  =  1+(u_{1,2}+1)x_1+(u_{1,2}+1)x_1(u_{4,2}+1)x_4,\\
\mathcal{F}_3  = 1+ (u_{2,3}+1)x_2, \\
\mathcal{F}_4  = x_3,\\
\mathcal{F}_5  = 1+ (u_{4,5}+1)x_4. 
\end{array}
\end{displaymath} 
Evaluating at $\textbf{y}_0=01111$ we obtain
\begin{displaymath}
\begin{array}{l}
\mathcal{F}_1  = 0, \ \  
\mathcal{F}_2  =  1, \ \
\mathcal{F}_3  = u_{2,3}, \ \
\mathcal{F}_4  = 1, \ \
\mathcal{F}_5  = u_{4,5}. 
\end{array}
\end{displaymath} 
Then, 01111 will be a steady state if and only if $\mathcal{F}_i=\textbf{y}_{0i}$ for $i=1,\ldots,5$. This gives the following solution $u_{2,3}=1$, $u_{4,5}=1$. Thus, in this case there is a unique control, $u_{2,3}=u_{4,5}=1$,  that guarantees $\textbf{y}_0=01111$ is a steady state (this control policy is illustrated in Figure~\ref{fig:summary}F, $u=D$). In general, the equations can be more complex and nonlinear, so computational algebra is needed. See the results section for applications to more complex models. 

\textit{Example }\ref{eg:toy} (cont.) We now show how the problem of creating new steady states can also be solved using node controls. We again use the toy network, and assume that we want to make $\textbf{y}_0=11110$ a steady state. For simplicity we only consider control of nodes $x_1$ (constant expression), $x_3$ (knock-out and constant expression), and node $x_4$ (constant expression). The network is
\begin{displaymath}
\begin{array}{l}
\mathcal{F}_1  = (u_1^+ + 1)(1+x_3+x_5+x_3x_5) + u_1^+,\\
\mathcal{F}_2  = 1+x_1+x_1x_4,\\
\mathcal{F}_3  = (u_3^-+u_3^+ + 1)(1+ x_2) + u_3^+, \\
\mathcal{F}_4  = (u_4^+ + 1)x_3 + u_4^+,\\
\mathcal{F}_5  = 1+ x_4. 
\end{array}
\end{displaymath} 
Evaluating at $\textbf{y}_0=11110$ we obtain
\begin{displaymath}
\begin{array}{l}
\mathcal{F}_1  = u_1^+,\ \
\mathcal{F}_2  = 1,\ \
\mathcal{F}_3  = u_3^+, \ \
\mathcal{F}_4  = 1,\ \
\mathcal{F}_5  = 0. 
\end{array}
\end{displaymath} 
Then, 11110 will be a steady state if and only if $u_1^+=1$ and $u_3^+=1$. That is, neither control by itself is sufficient, but together they create the steady state (this control policy is illustrated in Figure~\ref{fig:summary}F, $u=C$).

\subsubsection{Destroying existing steady states, or, in general, blocking transitions} 
\label{sec:blocking_transitions}
Suppose that $\textbf{x}_0\in \mathbb F_2^n$ is an undesirable steady state of $\mathbf{ F}(x)$, that is, $\mathbf{ F}(\textbf{x}_0) = \textbf{x}_0$ (for instance, it could represent a tumor proliferative cell state that needs to be avoided). The goal here is to find a set of controls such that $\mathcal{F}(\textbf{x}_0,u) \neq \textbf{x}_0$. More generally, we may want to avoid a transition between two states $\mathbf{x}_0$ and $\mathbf{z}_0$. That is, we want to find controls such that $\mathcal{F}(\textbf{x}_0,u) \neq \textbf{z}_0$.  To solve this problem consider the following equation,
\begin{equation}
\label{Eq:blocking}
( \mathcal{F}_1(\mathbf{x}_0,u)-\mathbf{z}_{01} +1 )\ldots
( \mathcal{F}_n(\mathbf{x}_0,u)-\mathbf{z}_{0n} +1 )=0.
\end{equation}
Equation~\ref{Eq:blocking} defines a polynomial equation in the $u$ parameters.
It can be shown that $\mathcal{F}(\textbf{x}_0,u) \neq \textbf{z}_0$ if and only if Equation~\ref{Eq:blocking} is satisfied.

\textit{Example }\ref{eg:toy} (cont.) Here we focus on finding edges to block the transition from $\textbf{x}_0=00101$ to $\textbf{F}(\textbf{x}_0)=01111$. For simplicity we only consider control using the edges 
$x_3\rightarrow x_1, 
x_5\rightarrow x_1,
x_2\rightarrow x_3,
x_3\rightarrow x_4$. 
The network is 
\begin{displaymath}
\begin{array}{l}
\mathcal{F}_1  = 1+(u_{3,1}+1)x_3+(u_{5,1}+1)x_5+\\  
\quad \quad \quad (u_{3,1}+1)x_3(u_{5,1}+1)x_5,\\
\mathcal{F}_2  =  1+x_1+x_1x_4,\\
\mathcal{F}_3  = 1+ (u_{2,3}+1)x_2, \\
\mathcal{F}_4  = (u_{3,4}+1)x_3,\\
\mathcal{F}_5  = 1+ x_4. 
\end{array}
\end{displaymath}

Evaluating at $\textbf{x}_0=00101$ we obtain
\begin{displaymath}
\begin{array}{lllll}
\mathcal{F}_1  = u_{3,1}u_{5,1}, 
\mathcal{F}_2  =  1,
\mathcal{F}_3  = 1, 
\mathcal{F}_4  = u_{3,4},
\mathcal{F}_5  = 1. 
\end{array}
\end{displaymath}
Then, Eq.~\ref{Eq:blocking} becomes 
$(u_{3,1}u_{5,1}+1)(u_{3,4}+1)=0$ which has two solutions: $u_{3,4}=1$ and $u_{3,1}=u_{5,1}=1$ (first control policy illustrated in Figure~\ref{fig:summary}F, $u=B$). The second solution is what we refer to as a combinatorial control policy. Neither $u_{3,1}=1$ or $u_{5,1}=1$ is sufficient, but combined they  block the transition. In general, the equations can be more complex and nonlinear, so computational algebra is needed.

\subsubsection{Blocking regions in the state space} 
\label{sec:blocking_all}
We now consider the case where we want the dynamics to avoid certain regions. For example, if a particular value of a 
variable, $x_k=a\in \mathbb{F}_2$ triggers an undesirable pathway, or is the signature of an abnormal cell, then we want all 
steady states of the system to satisfy $x_k\neq a$. In this case, we consider the systems of equations
\begin{equation}
\label{Eq:blocking_all}
\begin{split}
\mathcal{F}_j(x,u)- x_j=0 , j = 1,\dots,m,\\
x_k-a=0.
\end{split}
\end{equation}
Note that, in contrast to Sections \ref{sec:find_edge_del} and \ref{sec:blocking_transitions}, we are now using variables for $x$ instead of specific values. Since the steady states with $x_k=a$ are to be avoided, we want to find controls $u$ for which Equation \ref{Eq:blocking_all} has no solution.

\textit{Example }\ref{eg:toy} (cont.) Here we consider the toy network and focus on controlling nodes to avoid regions of the form $x_3=0$. For simplicity we only consider control using the nodes $x_2$ (knock-out), $x_3$ (constant expression), $x_4$ (constant expression). The network is
\begin{displaymath}
\begin{array}{l}
\mathcal{F}_1  = 1+x_3+x_5+x_3x_5,\\
\mathcal{F}_2  = (u_2^- + 1)(1+x_1+x_1x_4),\\
\mathcal{F}_3  = (u_3^+ + 1)(1+ x_2) + u_3^+, \\
\mathcal{F}_4  = (u_4^+ + 1)x_3 + u_4^+,\\
\mathcal{F}_5  = 1+ x_4. 
\end{array}
\end{displaymath} 
Then, Eq.~\ref{Eq:blocking_all} becomes 
\begin{displaymath}
\begin{array}{l}
 1+x_3+x_5+x_3x_5 +x_1=0,\\
(u_2^- + 1)(1+x_1+x_1x_4) +x_2=0,\\
(u_3^+ + 1)(1+ x_2) + u_3^+ + x_3=0, \\
 (u_4^+ + 1)x_3 + u_4^+ + x_4=0,\\
 1+ x_4+x_5=0,\\
 x_3=0.
\end{array}
\end{displaymath} 
In contrast with the previous examples, this system of equations cannot be analyzed by hand. In Section~\ref{sec:methods_solving} we will show how computational algebra gives an equivalent, but simpler, system of equations.

\subsection{Identifying control targets} 
\label{sec:methods_solving}
In each case of Section \ref{sec:methods_equations} we obtained a system of equations (or a single equation) that we need to solve to find the appropriate controls. This can be done using computational algebra tools. For instance,
we can compute the Gr\"obner basis 
of the ideal associated to Equation~\ref{eq:forceFP}, see~\cite{cox1998using},
\begin{equation}
\label{}
I=\left\langle \mathcal{F}_1(\textbf{y}_0,u)- y_{01},\ldots,\mathcal{F}_n(\textbf{y}_0,u)- y_{0n}\right\rangle.
\end{equation} 

\textit{Example }\ref{eg:toy} (cont.) Now we continue the previous example where the goal was to avoid regions of the form $x_3=0$. We encode the system of equations as 
$
I =\langle 
 1+x_3+x_5+x_3x_5 +x_1,
(u_2^- + 1)(1+x_1+x_1x_4) +x_2,
(u_3^+ + 1)(1+ x_2) + u_3^+ + x_3,
 (u_4^+ + 1)x_3 + u_4^+ + x_4,\\
 1+ x_4+x_5,
 x_3\rangle
$. Using computational algebra we find a Gr\"obner basis of this ideal:\\
$I=\langle
 u_3^+, u_2^-, x_5+u_4^+ + 1, x_4+u_4^+, x_3, x_2+1, x_1+u_4^+ \rangle$.
 
 Thus, the original system of equations has the same solutions as the system 
\begin{displaymath}
\begin{array}{lllll}
u_3^+=0,& u_2^-=0,& x_5+u_4^+ + 1=0,& x_4+u_4^+=0,
\\ x_3=0,& x_2+1=0,& x_1+u_4^+=0.
\end{array}
\end{displaymath} 

In order to avoid regions of the form $x_3=0$, we need to find parameters for which the system has no solutions. This is guaranteed if any of the polynomials is equal to 1. Namely, we select the equations that only have the control parameters $u_3^+=0, u_2^-=0$. If we use $u_3^+=1$ or $u_2^-=1$, the system is guaranteed to have no solutions. Thus, $u_3^+=1$ or $u_2^-=1$ independently are sufficient to achieve the control of the system (second control policy illustrated in Figure~\ref{fig:summary}F, $u=A$).

As we will show in Examples~\ref{p53_cancer_dna_damage}-\ref{T_LGL_network}, the computation of a Gr\"obner basis allows us to read out all controls for which the system of equations has a solution. Furthermore, our algebraic approach can detect combinatorial control actions (control by the synergistic combination of more than one action).
\section*{Results}
We test our control methods using two published models where potential intervention targets
were identified along with experimental validations. In the first example we focus on control with edge deletions while for the second we use control with node deletions and constant expressions.
\begin{example}\label{p53_cancer_dna_damage} \textbf{P53-mdm2 network.}
A Boolean network model for the signaling response of DNA damage in a \textit{p53} network
was developed in~\cite{Choi2012}; the wiring diagram of this model is reproduced in Figure~\ref{fig:p53_network}. 
The tumor suppressor protein \textit{p53} can trigger either cell cycle arrest or apoptosis in response to DNA damage
in normal cells. The authors of~\cite{Choi2012} extended their analysis to a breast cancer cell line, MCF7, with the purpose of
identifying potential therapeutic interventions in the network for the cancer cell.
Thus, in this example we will focus on the cancer cell model where \textit{PTEN} and \textit{p14ARf} are always inactive (fixed to zero) and
\textit{cyclinG} is always active (fixed to 1), see Table~5 of~\cite{Choi2012}.
This system can be represented as a discrete dynamical system $\mathbf{F}=(f_1,\ldots,f_{16}):\mathbb F_2^{16}\rightarrow \mathbb F_2^{16}$
with 16 nodes and 50 edges. We represent the nodes by
\begin{equation}\label{Eq:p43_node_labels}
\begin{tabular}{ll}
$x_1 = $ \textit{ATM}, & $x_2 = $ \textit{p53}, \\
$x_3 = $ \textit{Mdm2}, &$x_4 = $ \textit{MdmX}, \\
$x_5 = $ \textit{Wip1}, &$x_6 = $ \textit{cyclinG},\\
 $x_7 = $ \textit{PTEN}, &$x_8 = $ p21, \\
 $x_9 = $ \textit{AKT}, &$x_{10} = $ \textit{cyclinE}, \\
 $x_{11} = $ \textit{Rb}, &$x_{12} = $ \textit{E2F1}, \\
 $x_{13} = $ \textit{p14ARf}, &$x_{14} = $ \textit{Bcl2}, \\
 $x_{15} = $ \textit{Bax}, &$x_{16} = $ \textit{caspase}. \\
\end{tabular}
\end{equation}
\begin{figure}[!h]
\begin{center}
\includegraphics[width=3in]{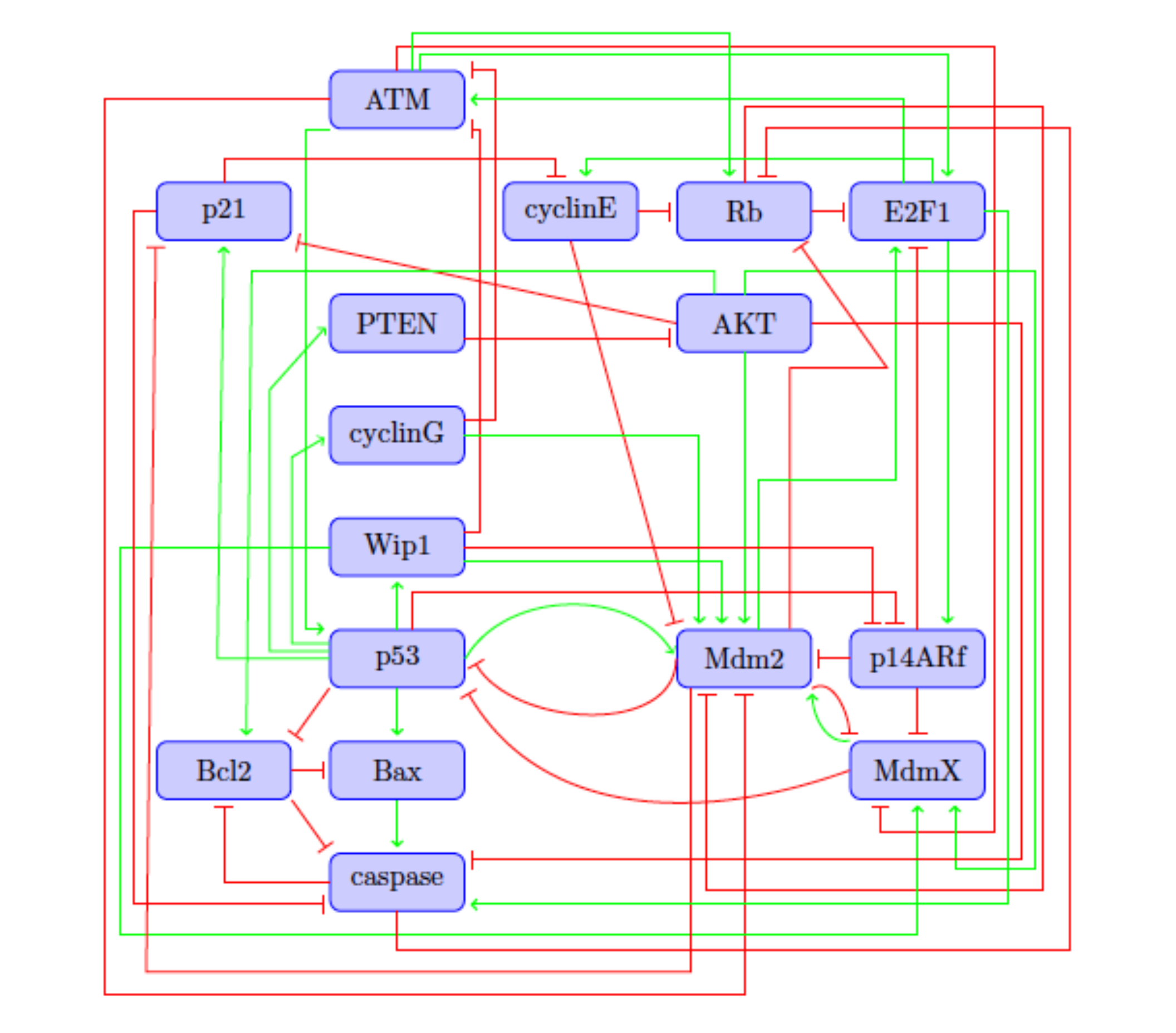}
\caption{The \textit{p53-mdm2} network adapted from~\cite{Choi2012}.
Arrows in green represent activation while hammerhead arrows (in red) represent inhibition. Self loops were omitted, see text in Example~\ref{p53_cancer_dna_damage} for an explanation.
For the cancer cell model, \textit{PTEN} and \textit{p14ARf} are always inactive (fixed to zero) and
\textit{cyclinG} is always active (fixed to 1).}
\label{fig:p53_network}
\end{center}
\end{figure}  

The polynomial functions for this network are listed in the supplementary material.
We remark that the original model in~\cite{Choi2012} considers threshold functions for all the regulatory rules and this type of functions might introduce
self-loops in the wiring diagram. That is, when we translated the threshold rule for $x_i$ into a polynomial $f_i$, the function $f_i$ might depend on $x_i$. 
Notice that the self-loops were omitted in Figure~\ref{fig:p53_network} to be consistent with the original model.

For this model, in the presence of DNA damage, the system has a single limit cycle representing the state of cell cycle arrest, where $p53$ and $p21$ are oscillating; see~Figure~\ref{fig:limit_cycle_p53}.

\begin{figure}[!h]
\begin{center}
\includegraphics[width=2.5in]{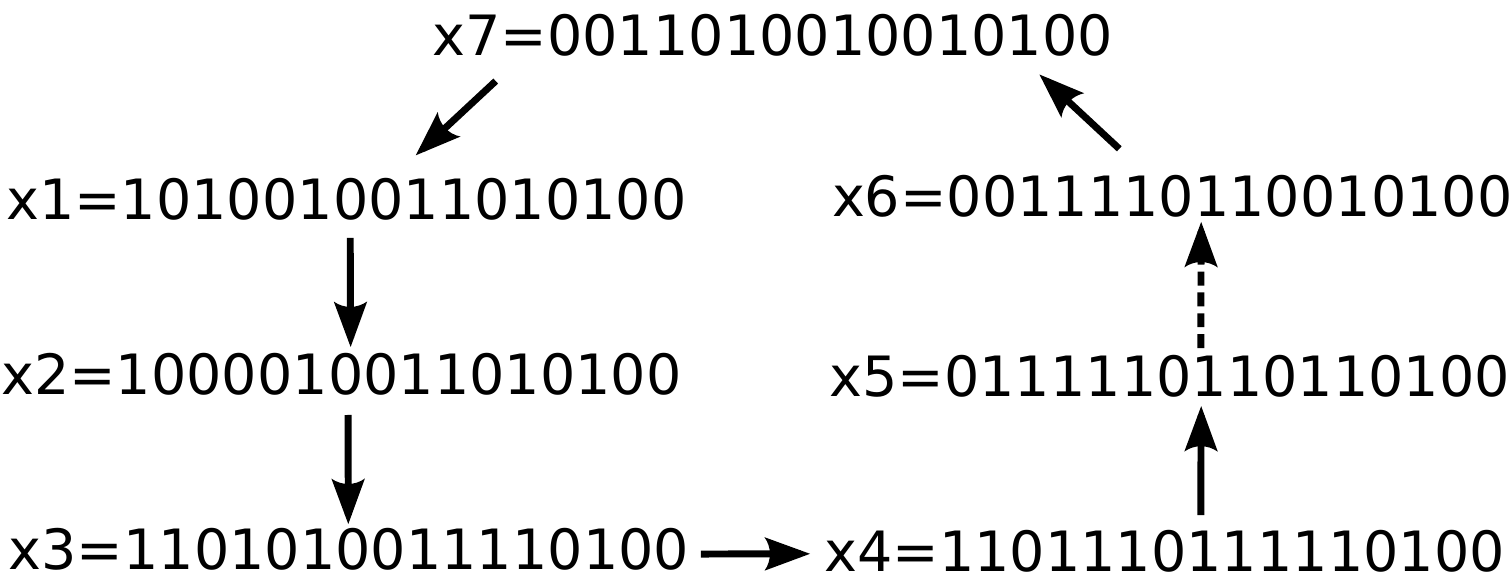}
\caption{States of limit cycle representing cell cycle arrest in the p53 model.
The order of the vector entries follows the indexing in Equation~\ref{Eq:p43_node_labels}.
The dashed edge represents the transition target to destroy the limit cycle. 
}
 \label{fig:limit_cycle_p53}
\end{center}
\end{figure}   

Suppose that we want to find out which set of edges one can manipulate in order to destroy
this limit cycle and to lead the system into a different attractor, one that represents a desirable cell state.
Let us start by considering a desirable state that represents cell death,
\begin{equation}
\label{p53_fixed_point}
\textbf{y}_0 = (1, 1, 1, 0, 0, 1, 0, 1, 1, 0, 0, 1, 0, 0, 1, 1),
\end{equation}
where $x_{16} = $ \textit{caspase} is active. In order to make $\textbf{y}_0$ a steady state with the approach described in the Methods section,
we form the following system of polynomial equations,
\begin{equation}
\label{p53_fixed_point_eval}
\mathcal{F}_i(\textbf{y}_0,\mu) = y_{0i}, j = 1,\dots,16.
\end{equation}
The solutions for the system of equations~\ref{p53_fixed_point_eval} are given by the nonzero generators
of the ideal associated to the system~\ref{p53_fixed_point_eval}, given in Equation~\ref{p53_gb_gen},
\begin{equation}
\label{p53_gb_gen}
\begin{tabular}{l}
$\{u_{2,5}+1, u_{16,11}(u_{1,11}+1),u_{8,8}(u_{3,8}+1),u_{2,2}(u_{3,2}+1),$\\
$u_{1,2}(u_{3,2}+1), u_{1,1}u_{12,1},u_{12,16}u_{16,16}(u_{8,16}+1)\}$\\
\end{tabular}
\end{equation}
There are $945$ solutions for Equation~\ref{p53_gb_gen}. Each solution gives a controller that
guarantees the desired steady state, but each controller may have a different impact on the dynamics of the network.
Table~\ref{tab:comparison_with_choi} (middle row) shows one of the controllers from Equation~\ref{p53_gb_gen}. 
In Table~S1 of the supplementary material, we list the top ten control combinations from Equation~\ref{p53_gb_gen}, that is, those that give the largest basin of 
$\textbf{y}_0$ as well as the ten sets that give the smallest basin for $\textbf{y}_0$; see~Table~S2.
For each solution in tables~S1-S2, we crossed-out the edges that become nonessential after the other controllers in the set have been applied;
see the discussion for more about nonessential edges.

Furthermore, we can aim to destroy the limit cycle in~Figure~\ref{fig:limit_cycle_p53}.
We target one of the transitions within this limit cycle,
let us take the transition $\textbf{x}_5\rightarrow\textbf{x}_6$, see dashed transition in Figure~\ref{fig:limit_cycle_p53}. 
Blocking this transition gives additional controllers that help to stabilize the system at the desired fixed point;
see the last row of Table~\ref{tab:comparison_with_choi}.

The deletion of the edges $\textit{mdm2}\rightarrow\textit{p53}$ and $\textit{p53}\rightarrow\textit{Wip1}$
were identified in~\cite{Choi2012} by deleting one edge at a time and checking if the modified system had the desired attractor.
\cite{Choi2012} reported that the combinatorial action of these controllers increased the basin of attraction of the desired fixed point to more than 50\%,
which was validated experimentally.
However, doing this type of search for finding all possible combinations of controls is infeasible.
Since for each edge we have 2 possible actions (control or no control), and there are
50 edges, then there are $2^{50}$ networks to be analyzed in total. 
In contrast, the computational algebra approach of this paper allows to obtain all combinations of edges that guarantee the desired 
steady state of the system in one process.
\begin{table}
\caption{Difference of impact in the combinatorial action of edge deletions.
Control edges that increase the basin of attraction of cell death represented by $\textbf{y}_0$ in Equation~\ref{p53_fixed_point}.
There are $2^{16}=65536$ possible states. The number in parentheses is the ratio between the basin size and the total number of states.}
\label{tab:comparison_with_choi}
  \centering
\begin{tabular}{| l | l | l |}
\hline
Controllers applied & \multicolumn{1}{c|}{Ref.} &Basin size of $\textbf{y}_0$\\ \hline
$\textit{mdm2}\rightarrow\textit{p53}$&\multirow{2}{*}{\cite{Choi2012}}&\multirow{2}{*}{35581 (54.29\%)}\\
$\textit{p53}\rightarrow\textit{Wip1}$ & & \\ \hline
$ p53\rightarrow Wip1$  & A control set that & \multirow{4}{*}{39856 (60.82\%)} \\ 
$ Mdm2\rightarrow p21$& forces $\textbf{y}_0$ to be& \\
$ Mdm2\rightarrow p53$ & a fixed point, from& \\
$ p21 \rightarrow Caspase $& Equation~\ref{p53_gb_gen}.& \\ \hline 
$\textit{mdm2}\rightarrow\textit{p53}$& A control set to make &\multirow{9}{*}{65536 (100\%)}\\
$\textit{p53}\rightarrow\textit{Wip1}$ &  $\textbf{y}_0$ a fixed point &\\
$\textit{mdm2}\rightarrow\textit{p21}$&  and for blocking &\\ 
$\textit{p21}\rightarrow\textit{Caspase}$& the transition in red &\\ 
$\textit{ATM}\rightarrow\textit{Rb}$& at Figure~\ref{fig:limit_cycle_p53}.&\\ 
$\textit{mdm2}\rightarrow\textit{Rb}$&  &\\ 
$\textit{mdmx}\rightarrow\textit{p53}$&  &\\ 
$\textit{Rb}\rightarrow\textit{E2F1}$&  &\\ 
$\textit{Bcl2}\rightarrow\textit{Bax}$&  &\\ 
\hline
\end{tabular}
\end{table}
\end{example}
\begin{example}\label{T_LGL_network}  \textbf{T-LGL network.}
A Boolean network model for the blood cancer large granular lymphocyte (T-LGL) leukemia 
was developed in~\cite{Zhang:2008aa} and further analyzed in~\cite{Saadatpour:2011aa, Zanudo:2015aa}.
T-LGL leukemia is characterized by escaping cell death through abnormal mechanisms,
which are insensitive to \textit{Fas}-induced apoptosis,~\cite{Zhang:2008aa}.
This network has 60 nodes but was reduced to a  subnetwork of 16 nodes for steady state analysis, see Figure~\ref{fig:tlgl_network}.
Here we use the 16-nodes network to identify potential control targets.

\begin{figure}[!h]
\begin{center}
\includegraphics[width=3in]{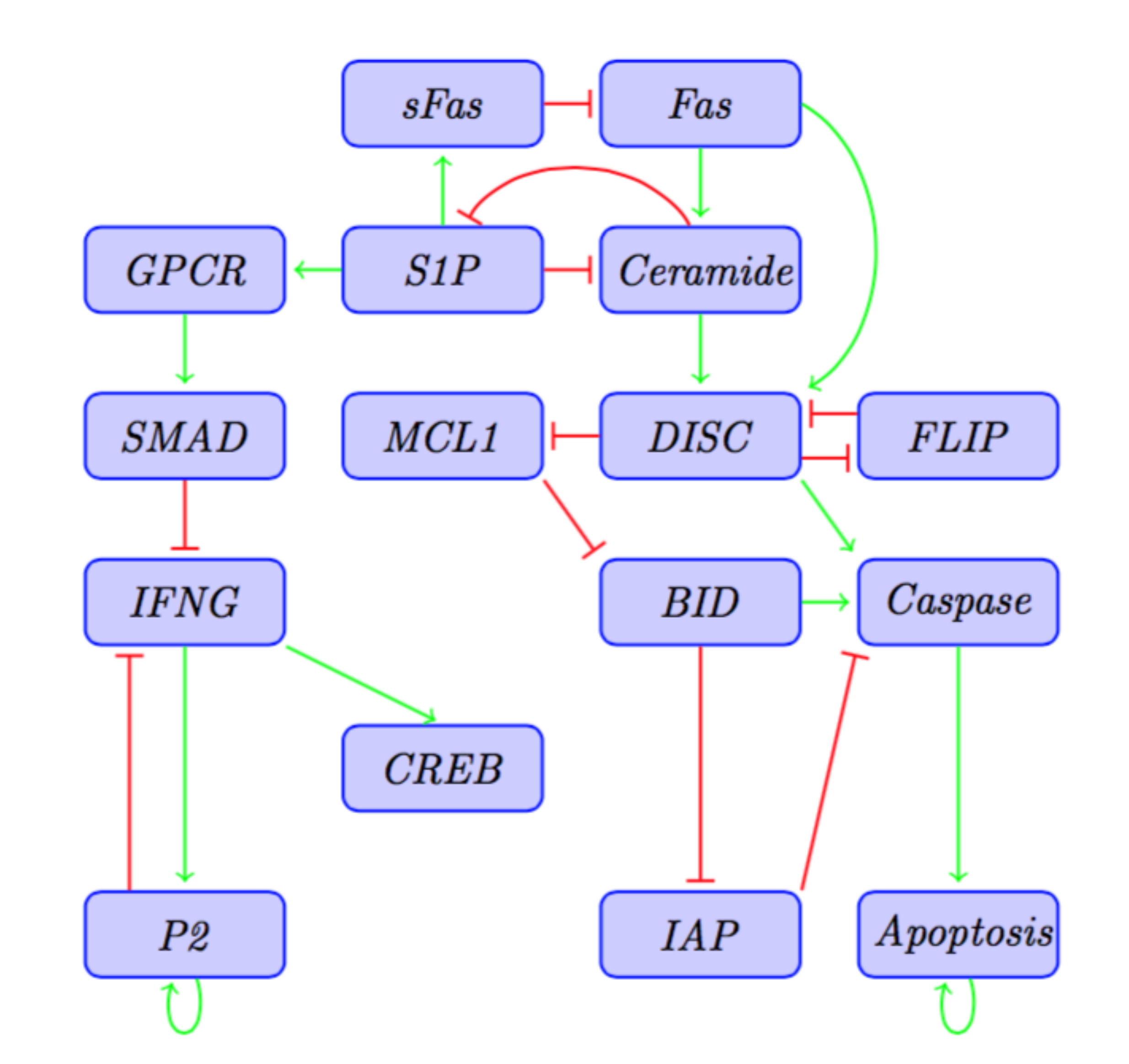}
\caption{Reduced \textit{T-LGL} network adapted from~\cite{Saadatpour:2011aa}.
Arrows in green represent activation while hammerhead arrows (in red) represent inhibition.
All the negative edges from Apoptosis were omitted for simplicity.}
\label{fig:tlgl_network}
\end{center}
\end{figure}

We represent the nodes by
\begin{center}
\begin{tabular}{lll}
$x_1=$\textit{CREB}, & $x_2=$\textit{IFNG}, & $x_3=$\textit{P2},\\
$x_4=$\textit{GPCR}, & $x_5=$\textit{SMAD}, & $x_6=$\textit{Fas},\\
$x_7=$\textit{sFas}, & $x_8=$\textit{Ceramide}, & $x_9=$\textit{DISC},\\
$x_{10}=$\textit{Caspase}, & $x_{11}=$\textit{FLIP}, & $x_{12}=$\textit{BID},\\
$x_{13}=$\textit{IAP}, & $x_{14}=$\textit{MCL1}, & $x_{15}=$\textit{S1P},\\
$x_{16}=$\textit{Apoptosis}. & &\\
\end{tabular}
\end{center}
The polynomial rules for this system are listed in Equation~\ref{Eq:T-GLG-polynomials},
\begin{equation}\label{Eq:T-GLG-polynomials}
\begin{tabular}{l}
$f_1=x_2(1+x_{16})$, $f_2=(x_5+1)(x_3+1)(1+x_{16})$,\\
$f_3= (x_2x_3+x_2+x_3)(1+x_{16})$,\\
$f_4= x_{15}(1+x_{16})$, $f_5= x_4(1+x_{16})$,\\
$f_6= (x_7+1)(1+x_{16})$, $f_7= x_{15}(1+x_{16})$,\\
$f_8= (x_{15}+1)x_6(1+x_{16})$,\\
$f_9= (x_6x_8x_{11}+x_6x_8+x_6x_{11}+x_6+x_8)(1+x_{16})$,\\
$f_{10}= (x_9x_{12}x_{13}+x_9x_{12}+x_{12}x_{13}+x_9+x_{12})(1+x_{16})$,\\
$f_{11}= (x_9+1)(1+x_{16})$, $f_{12}= (x_{14}+1)(1+x_{16})$,\\
$f_{13}= (x_{12}+1)(1+x_{16})$, $f_{14}= (x_9+1)(1+x_{16})$,\\
$f_{15}= (x_8+1)(1+x_{16})$,\\
$f_{16}= x_{10}x_{16}+x_{10}+x_{16}$.
\end{tabular}
\end{equation}
This system has three steady states, one that represents the normal state,
$\textbf{x}_0 =0000000000000001$, where \textit{Apoptosis} is ON,
and 0001101000101110, \\
0011101000101110, the disease states in which \textit{Caspase} and \textit{Apoptosis} are OFF.

To find the potential node deletions (or constant expressions) that will block the disease states we use Eq. \ref{Eq:blocking_all},
\begin{equation}
\label{Eq:blocking_fixed_point_ex}
\begin{split}
(u^{-}_{1}+u^{+}_{1}+1)f_{1}(\textbf{x})+u^{+}_{1} - x_1=0\\
\vdots\hspace*{0.15in}\\
(u^{-}_{16}+u^{+}_{16}+1)f_{16}(\textbf{x})+u^{+}_{16} - x_{16}=0\\
x_{16}=0, x_{10}=0
\end{split}
\end{equation}
Equation~\ref{Eq:blocking_fixed_point_ex} encodes all parameters for which there is a steady state where \textit{Caspase} and \textit{Apoptosis} are OFF. Thus, we are interested in the parameters for which this system of equations has no solution. Since for each node we have 3 possible actions (no control, node deletion, constant expression), there are $3^{16}$ networks to be analyzed in total. Thus, even for this small network, exhaustive search is computationally challenging.

On the other hand, computational algebra allows us to obtain the parameter combinations that guarantee that the disease states are \emph{not} fixed points of the system (see the supplementary material for details). 
The parameter combinations (enclosed in brackets, where entries not shown are equal to zero) are
\begin{equation}
\label{Eq:blocking_fixed_point_ex_eval}
\begin{split}
 \{u^+_8=1\}, \{u^+_{9}=1\}, \{u^+_{10}=1\},  \{u^+_{12}=1\}, \{u^-_{14}=1\},\\
  \{u^-_{15}=1\}, \{u^+_{6}=1,u^-_{11}=1\},  \{u^-_{7}=1,u^-_{11}=1\}.
\end{split}
\end{equation}

Thus we obtain that the constant expression of \textit{Ceramide}, \textit{DISC}, \textit{Caspase}, or  \textit{BID}, or the deletion of \textit{MCL1} or \textit{S1P}, will guarantee that the disease states are not steady states of the system (Table~\ref{tab:control_nodes_T_LGL}). These controls could also be found by trying one control at a time as was done in \cite{Saadatpour:2011aa}. Importantly, our computational algebra approach shows that there are two additional control policies that consist of a combination of different controls. It can be shown that neither 
\textit{Fas}, \textit{sFas}, \textit{FLIP} individually can eliminate the disease states, but the deletion of \textit{FLIP} \textbf{combined} with the constant expression of \textit{Fas} or the deletion of \textit{sFas} will work (Table~\ref{tab:control_nodes_T_LGL}). Furthermore, those are the only combinations that guarantee the disease states will not be attractors.

We note that, in the worst case, computing the Gr\"obner basis for a system of polynomial equations has doubly exponential complexity in the number of solutions.
However, for the type of networks discussed in this paper, namely, biological networks where most of the nodes are regulated by only a small subset of the other nodes, Gr\"obner bases can be computed in a reasonable time, see~\cite{Hinkelmann:2011aa}.
\newpage
\begin{table}[!h]
\caption{Control nodes for the reduced T-LGL network.
The last two rows represent combinatorial actions of two nodes.
All attractors are steady states, and the basin sizes include the steady states themselves.
Notice that node $x_{16} = $ \textit{Apoptosis} is a conceptual node in this model, thus it is not a relevant solution for network control.}
\label{tab:control_nodes_T_LGL}
\begin{tabular}{| c | l | c | c |}
\hline
Solution & Control targets & Attractor & Basin size\\ \hline
$u^+_8=1$ & \textit{Ceramide}=ON    & 0000000100000001 & 100\%  \\ \hline
$u^+_{9}=1$ & \textit{DISC}=ON    &   0000000010000001 & 100\% \\ \hline
$u^+_{10}=1$ & \textit{Caspase}=ON  & 0000000001000001 & 100\%  \\ \hline
$u^+_{12}=1$ & \textit{BID}=ON      & 0000000000010001 & 100\%   \\ \hline
$u^-_{14}=1$ & \textit{MCL1}=OFF    & 0000000000000001 & 100\%   \\ \hline
$u^-_{15}=1$ & \textit{S1P}=OFF     & 0000000000000001 & 100\%   \\ \hline
$u^+_{6}=1$ & \textit{Fas}=ON  & \multirow{2}{*}{0000010000000001} & \multirow{2}{*}{100\%}  \\
$u^-_{11}=1$ & \textit{FLIP}=OFF  & &\\ \hline
$u^-_{7}=1$ & \textit{sFas}=OFF  & \multirow{2}{*}{0000000000000001} & \multirow{2}{*}{100\%}  \\
$u^-_{11}=1$ & \textit{FLIP}=OFF  & &\\
\hline
\end{tabular}
\end{table}
\end{example}
\section*{Discussion}
The design of control policies for gene regulatory networks is an important challenge in systems biology.
The method described here exploits the interplay between the structure and the dynamics of the network to identify potential
control interventions that will drive the system towards desired dynamics. The formulation of the problem as that
of finding all solutions to a system of polynomial equations provides an alternative to exhaustive search of all possible
combinations of interventions, which often is not feasible. 

One shortcoming of the method is that, in its current form, it requires the Boolean network model to be updated synchronously,
with deterministic dynamics. While steady states of the network do not depend on the update order used, general 
limit cycles and attractor basins do, however. Thus, 
some of the methods described here might not be applicable in a stochastic setting. For instance, if the dynamics is generated
using the asynchronous update or more generally using the stochastic settings in~\cite{Murrugarra2012,Shmulevich2002,R.Thomas1991,Thomas:1990aa},
then encoding the controllers for blocking transitions will need to have a different setup than the one described here.
However, the method for producing a new steady state is still valid for all variants of stochastic updates because the system will maintain the steady state.
Another shortcoming is that the control methods described in this paper were developed for Boolean networks only. However, many published discrete models of GRNs are multistate. These methods could potentially be extended to a more general setting, where the network variables might attain more than two states
~\cite{Murrugarra:2012aa,Murrugarra:2011aa,Veliz-Cuba:2010aa,Thomas:1990aa,Ahmad:2012fj}. We remark that it is possible to map a multistate model into a Boolean model (see~\cite{Didier:2011kq}) where
our methods can be applied and then it would be possible to recover a control set for the multistate model.

An important challenge in the process of identification of control targets in a network is to develop
methods that can identify controllers that can guarantee global reachability of a
desired steady state. This problem is not addressed in this paper, and remains to be studied in the future.
For instance, the control strategies do not give any information about the basin size of a fixed point generated by the methods of this paper.
However, we remark that some algebraic methods allow to estimate the change in the basin size after an edge deletion, see~\cite{Murrugarra:2015uq}.
Nonetheless, the control targets identified by the algebraic techniques described here could be used for further analysis of the system,
such as for studying reachability~\cite{Li:2015aa}, or for designing optimal control policies in a stochastic setting~\cite{yousefi2012,Yousefi15072013,6557489,Yousefi:2014aa}. 

Finally, it is worth pointing out that the methods of this paper might produce a large number of control
strategies, which all give the desired result, but many of these might be biologically meaningless or
infeasible as actual interventions. Eliminating those might be challenging. For instance, some solution sets might contain
nonessential~\cite{Li:2013aa} or nonfunctional~\cite{Comet2013} edges. That is, an edge could become nonessential after the other controllers in the set have been applied. 
In Table~S1 of the supplementary material, we list the top ten control combinations for Example~\ref{p53_cancer_dna_damage}
where we crossed-out the edges that become nonessential after the other controllers have been applied. Moreover, one can
group the solution sets by considering only minimal sets where all the control edges are functional. For instance, we can group
the first four solutions of Table~S1 of the supplementary material into one group with the minimal representative set given in the middle row of Table~\ref{tab:comparison_with_choi}   
\section*{Conclusions}
This paper presents a novel approach to the identification of potential interventions in
Boolean molecular networks. The methods use the theoretical foundations of algebraic geometry to encode
the structure of a network by a set of polynomials and then, with the use of computer algebra techniques, 
find a set of nodes and edges to perform interventions {\it in silico}. The methods were validated using two published
models where dynamic network interventions were identified, the \textit{p53-mdm2} system and the T-LGL leukemia model.
It was shown that the methods in this paper can identify the controllers that were already reported and also find new potential targets.
Some of these new control targets are combinatorial in nature and might result in more efficient strategies as was shown in the results
section using the change in the basin size of the system as a measure of efficiency.
\begin{backmatter}
\section*{Competing interests}
The authors declare that they have no competing interests.

\section*{Author's contributions}
DM, AVC, BA, and RL conceived the study. DM and AVC performed the numerical experiments and theoretical analysis. BA designed code. All authors helped in the writing of the manuscript.
All authors approved the final version of the manuscript.
\section*{Acknowledgements}
The authors thank the referees for their very insightful comments that have improved the manuscript.
RL was supported by a grant from the U.S. Army Research Office, Grant Nr. W911NF-14-1-0486. 
\bibliographystyle{bmc-mathphys} 
\bibliography{ref_for_control}      

\end{backmatter}
\end{document}